# Change in the mass composition of primary cosmic radiation at energies in the range of $E_0 = 1\text{-}100$ PeV according to data of the KASCADE-Grande experiment

S. E. Pyatovsky[1]
A. D. Erlykin, V. S. Puchkov

P.N. Lebedev Physical Institute
The Russian Academy of Sciences, Russia

The change in the fluxes of the lightest and heaviest nuclei in the mass composition of primary cosmic radiation in the energy range of $E_0 = 1\text{-}100$ PeV is analyzed on the basis of data from the KASCADE-Grande experiment. This analysis is performed by means of the min-max age ($S_{\text{min-max}}$) method for extensive air showers (EAS) that is proposed in the present article. This method makes it possible to estimate the exponents of the spectra for the lightest and heaviest nuclei in the mass composition of primary cosmic radiation and to study irregularities in these spectra at energies $E_0$ in the range under consideration. The $S_{\text{min-max}}$ method for estimating the mass composition of primary cosmic radiation is based on a sizable sample (more than 100 million events) of data on EAS features in the range of $E_0 = 1\text{-}100$ PeV from the KASCADE-Grande experiment. It is shown that, in the primary-energy range of $E_0 = 1\text{-}100$ PeV, the exponent in the integrated spectrum changes from 2.1 to 2.7 for the lightest nuclei in the mass composition of primary cosmic radiation and from 1.5 to 2.1 for the heaviest nuclei. These results comply with the conclusion drawn by the KASCADE-Grande Collaboration that the knee in the $E_0$ spectrum of nuclei of primary cosmic radiation in the range of $E_0 = 3\text{-}5$ PeV is due to the elimination of light nuclei. The bump appearing in the $E_0$ spectrum of primary cosmic radiation according to the database of the KASCADE-Grande experiment and lying in the range of $E_0 = 50\text{-}75$ PeV is analyzed by the $S_{\text{min-max}}$ method and is found to agree with data of the GAMMA (GAMMA-07) experiment.

1. Introduction

The mass composition of primary cosmic radiation (PCR) has been studied both on the basis of features of events in the cores of extensive air showers (EAS) by the halo method characterized by a spatial resolution of about 30 $\mu$m and proposed for estimating the fraction of $p$ + He light nuclei in the PCR mass composition in [1] and on the basis of the distribution of individual-shower features, such as the number of charged particles in EAS ($N_e$), the number of muons in EAS ($N_\mu$), and the depth for the maximum of EAS development from the boundary of the atmosphere ($X_{\text{max}}$). The estimation of the PCR mass composition on the basis of EAS features depends substantially on the parameters of the model used to reconstruct EAS in the atmosphere. At the same time, a statistical set of gamma-ray families that are characterized by the presence of a halo, whose nature is explained in [2], and which are localized in the shower cores is used within the halo method for estimating the fraction of $p$ + He light nuclei in the PCR mass composition. These families carry information about the earliest events of the interaction of PCR nuclei with atoms of the Earth's atmosphere; therefore the halo method is weakly model-dependent. However, the halo method has the restriction of providing an estimate of the fraction of $p$ + He light nuclei in the PCR mass composition at a specific value of $E_0$. The result obtained in [3] by estimating the fraction of $p$ + He light nuclei in the PCR mass composition at $E_0 = 10$ PeV by the halo method exceeds (39 ± 6)%; otherwise, the halo would be unobservable in experiments with X-ray emulsion chambers (XREC). The halo method is based on substantially different probabilities for the formation of halos in EAS cores and for the detection of halos in XREC that are initiated by PCR nuclei belonging to different sorts. However, the halo method does not provide a statistically significant sample of events for constructing the $E_0$ spectrum of primary cosmic radiation, and this gives no way to estimate the change in the PCR mass composition for energies in the range of $E_0 = 1\text{-}100$ PeV, in which we are interested. The energies $E_0$ of nuclei belonging to showers whose cores are recorded as gamma-ray families in XREC are not measured directly either. However, these energies, which are model-dependent features of gamma-ray families, can be estimated either by employing the expression $E_0 = k\Sigma E_\gamma$, where $k \cong 10$ for EAS initiated by protons and $k \cong 70$ for EAS initiated by iron nuclei, or by considering the development of electron-photon cascades in lead XREC.

---

[1] vgsep@ya.ru

Table 1. Fractions of $p + $ He nuclei and nuclei heavier than helium in the PCR mass composition according to estimations by means of the halo method on the basis of data from the Γ-block of the X-ray emulsion chambers in the Pamir experiment near the bump of the $E_0$ spectrum of primary cosmic radiation for $E_0 = $ 3-5 PeV.

| $E_0$, PeV | $P_{p + \text{He}}$, % | $P_{>\text{He}}$, % |
|---|---|---|
| ≲ 3 | 67-100 | 0-33 |
| 3-5 | 58-78 | 22-42 |
| 5-10 | 45-67 | 33-55 |

The halo method was used to estimate the fractions of $p + $ He and nuclei heavier than helium in the PCR mass composition at the most probable halo formation energy of $E_0 = 10$ PeV, which was determined as an average weighted with probabilities for halo formation by various PCR nuclei. The values obtained in this way for the $p + $ He fraction and the fraction of nuclei heavier than helium in the PCR mass composition are listed in Table 1. The values in Table 1 show that, in the $E_0$ range extending up to 10 PeV, the fraction of $p + $ He light nuclei decreases, while the fraction of medium-heavy and heavy nuclei increases. As $E_0$ grows, the probability for halo formation by medium-heavy, heavy, and super heavy groups of nuclei in the PCR mass composition becomes higher, and the halo method ceases to be applicable at high $E_0$, since the halo signature turns out to be insensitive to sorts of nuclei that initiated EAS whose cores are recorded in XREC as halos.

Samples of events are substantially vaster in experiments studying EAS than in experiments employing XREC. However, sizable fluctuations and a model dependence of features of individual air showers lead to a substantial spread of the estimates of fractions of nuclei in the PCR mass composition at energies in the range of $E_0 = $ 1-100 PeV. For example, the estimations of the fractions of protons and helium nuclei in the PCR mass composition in different experiments differ in this $E_0$ range by 5 to 50% [1]. At the same time, the change in the fraction of the lightest and heaviest nuclei in the PCR mass composition can be estimated not only on the basis of EAS features obtained experimentally but also on the basis of EAS statistics. The number of showers detected in an experiment should be sufficient for EAS features that could be classed with the ones typical of a given PCR nucleus, such as the EAS age $S$, to be able to manifest themselves in overall statistics.

The KASCADE-Grande experiment (Karlsruhe, Germany) numbers among experiments that collected a vast statistical sample of showers, with their features such as $N_e$, $N_\mu$, and $E_0$. This experiment provided open access to its data [4] for analysis, which is still under way. The experimental setup is deployed at a depth of about 1000 g/cm$^2$ of the standard atmosphere and is intended for EAS studies at energies of PCR nuclei in the range of $E_0 = 0.1$ PeV – 1 EeV. In the KASCADE-Grande experiment, the fraction of $p + $ He light nuclei in the PCR mass composition at energies in the range of $E_0 = $ 3 PeV – 0.3 EeV was estimated at 10 to 30%, with a knee in the spectrum of PCR light nuclei at $E_0 = 3$ PeV and a dominance of helium nuclei being observed [5].

The proposed age min-max ($S_{\text{min-max}}$) method is based of significant statistics of EAS features accumulated in the KASCADE-Grande experiment, which include EAS parameters that make it possible to estimate the change in the fractions of the lightest and heaviest nuclei in the PCR mass composition at energies in the range of $E_0 = $ 1-100 PeV.

2. Estimating changes in the PCR mass composition on the basis of the EAS age

The investigations reported in [6] reveal that the EAS age determined from the lateral distribution function as a parameter of the Nishimura-Kamata-Greisen function characterizes the mass numbers ($A$) of PCR nuclei. For a given air-target depth, at which the experimental setup is deployed, the EAS age decreases with increasing $N_e$ in proportion to log$^{-1}N_e$ for a given sort of PCR nuclei and increases with increasing mass number of PCR nuclei at a given $N_e$. Depending on $S(N_e)$, there are $N_e$ ranges where, with increasing $N_e$, the EAS age increases, decreases, or remains unchanged. In the $N_e$ range where $S$ remains unchanged, its growth with increasing $A$ is counterbalanced by its reduction with increasing $N_e$. It is shown in [6] that showers having minimum values of $S$ are initiated by lighter nuclei, whereas showers having maximum values of $S$ are initiated by heavier nuclei in the PCR mass composition. The sensitivity of $S$ to variations in the PCR mass composition follows, for example, from data of the HAWC experiment [7], which employed a setup deployed at a depth of 630 g/cm$^2$ of the standard atmosphere and which showed the reduction of $S$ with increasing $E_0$. The data of the ARGO-YBJ experiment (610 g/cm$^2$) [8] led to the dependence $S(X_{max})$, from which it follows that the change in $S$ as a function of the depth of the maximum of EAS development in the atmosphere, $X_{max}$, is independent of the sort of PCR nuclei, but that lighter nuclei of primary cosmic radiation correspond with a higher probability to minimum values of $S$ and larger values of $X_{max}$, while heavier nuclei correspond to smaller values of $X_{max}$ and larger values of $S$:

$$S = -(313 \pm 5) \times 10^{-5} X_{max} \text{ [g/cm}^2\text{]} + (2.94 \pm 0.02), \; R^2_a \cong 0.99 \quad (1)$$

According to data of the KASCADE-Grande experiment, the change in the EAS age with the depth of the experimental setup in the atmosphere can be estimated on the basis of the angular distribution of detected air showers:

$$S = -(0.092 \pm 0.005)(1 - \sec\theta) + (0.897 \pm 0.002), \; R^2_a \cong 0.97 \quad (2)$$

This shows that, for air-target depths in the range between 600 and 1000 g/cm$^2$, the growth of $S$ with the air-target depth is immaterial for the problem of estimating the change in the PCR mass composition by the $S_{min-max}$ method.

The EAS age $S$ was analyzed on the basis of about 100 million events recorded by the KASCADE-Grande experiment [1, 3]. This statistical sample of features of experimental EAS events is sufficient for the features of the showers formed by specific sorts of PCR nuclei to manifest themselves in the total data sample. The changes $\Delta S$, $\Delta \log N_e$, and $\Delta \log A$ were analyzed in order to estimate $<A>$. The results of investigations performed in [1-3, 5] show that, in the range of $\log N_e = 6.0$-6.5, the age undergoes virtually no change, $\Delta <S> \cong 0$. This is because the growth of $<S>$ with increasing $A$ counterbalances its reduction with increasing $N_e$. This suggests that the mass composition becomes heavier in this range of $N_e$. In the range of $\log N_e = 4.5$-5.0, $\Delta \log A \cong 0$, so that $<A>$ does not change. Under the assumption that $<A> = 4$ corresponds to $\log N_e = 4.5$, which follows from the analysis of experimental data on $S(N_e)$ in [1], it turns out that a group of PCR nuclei with the average mass number of $<A> = 12$, which is not larger than that for the CNO group, corresponds to $\log N_e \cong 6.0$, where the results of investigations performed at Moscow State University revealed for the first time an irregularity (knee) in the spectrum of primary cosmic radiation [9] in the energy range of $E_0 = 3$-5 PeV.

Similar estimations for the PCR mass composition were also performed on the basis of the muon component of EAS and by means of the halo method [1]. These investigations reveal that, in the energy range of $E_0 = 1$-100 PeV, the PCR mass composition undergoes a shift toward higher masses, but the average mass number $<A>$ of PCR nuclei does not become greater than that for the CNO group, the minimum fraction of light nuclei in the PCR mass composition at $E_0 = 10$ PeV being about 40%. According to the estimation on the basis of the ratio $N_\mu/E_0$, the average value of the mass number of PCR nuclei is $<A> = 5$-17 in the energy range of $E_0 = 2$-35 PeV. This complies with the results of the analysis of data from other experiments – for example, the TALE experiment at the Pierre Auger Observatory, where $<\ln A> = 2$ at $E_0 = 2$ PeV [10]. The average mass number $<A>$ changes both because of the increase in the flux of the heaviest nuclei in the PCR mass composition and because of the decrease in the flux of the lightest nuclei – possibly protons.

The $<A>$ values estimated on the basis of $S$ and $N_\mu$ in the range of $E_0 = 2$-35 PeV agree within the errors – that is, the relative growth of the flux of He-CNO nuclei should compensate for the relative growth of the flux of the heaviest nuclei in the PCR mass composition under conditions of the reduction of the relative flux of the lightest nuclei. This change in the PCR mass composition is possible if there are two local sources in the region between the knee and the boundary at $E_0 = 100$ PeV. The first source is responsible for the appearance of He-CNO nuclei in the PCR mass composition, while the second one provides the injection of the heaviest nuclei – for example, from supernovae.

2.1. Method of $S_{min-max}$ for estimating the change in the PCR mass composition

The $S_{min-max}$ method is based on employing statistics of EAS features that are sensitive to the PCR mass composition. The $S_{min-max}$ method ensures the formation of statistics of the lightest and statistics of the heaviest nuclei in the PCR mass composition.

The KASCADE-Grande database [4], which contains information about features of more than 150 million air showers, is used in the calculations. Among EAS features, the KASCADE-Grande database includes the EAS age $S$ recorded in the experiment over the range between 0.10 and 1.48; the energies $E_0$ that the nuclei initiating EAS have and which the experiment determined in the range of $\log(E_0 \text{ [eV]}) = 14.00\text{-}18.00$; and the numbers of electrons detected in the range of $\log N_e = 1.00\text{-}8.70$, $N_e$, and the number of muons detected in the range of $\log N_\mu = 1.00\text{-}7.70$, $N_\mu$, in nuclear-electromagnetic cascades. In particular, EAS features such as the zenith (0°-60°) and azimuthal angles of EAS arrival, the EAS detection time, the atmospheric pressure and temperature at the instant of EAS detection, and coordinates of the EAS axis can be analyzed with the aid of the KASCADE-Grande database.

From the results obtained by analyzing the sensitivity of the EAS age to the sort of the primary nucleus, it follows that the database of the statistical samples of EAS features that is arranged in the order of increasing $S$ contains, at its beginning, parameters of EAS initiated by the lightest nuclei in the PCR mass composition and, at its end, parameters of EAS initiated by the heaviest nuclei in the PCR mass composition that were recorded in the experiment. As the statistical sample of air showers becomes vaster (above 150 million events), the first (or last) EAS in age are initiated with a growing probability by the lightest (heaviest) nuclei in the PCR mass composition. Within the $S_{\text{min-max}}$ method, use is made of the database of EAS features that is sorted in the EAS age parameter rather than in the EAS detection time. In the overall statistical sample of EAS detected in the experiment, the first air showers in $S$ [in the order of increasing $S$ (lightest nuclei) or in the order of decreasing $S$ (heaviest nuclei)] are chosen, and their characteristics are studied.

The $S_{\text{min-max}}$ method for estimating the change in the PCR mass composition upon a change in $E_0$ has natural restrictions in studying the resulting spectra of groups of nuclei. In selecting EAS by the $S_{\text{min-max}}$ method, it is impossible to pinpoint the sort of the nucleus that initiated a given shower. However, one can state that the selected group of EAS includes showers formed by the lightest or the heaviest nuclei in the PCR mass composition at a given value of the energy $E_0$ of nuclei. The absolute normalization of the fluxes of various sorts of nuclei is not known either, since equal numbers of the lightest, heaviest, and mixed-composition nuclei of primary cosmic radiation are selected for analysis. Therefore, the percentage of the lightest and heaviest nuclei in the PCR mass composition cannot be assessed. Also, the $S_{\text{min-max}}$ method gives no way to estimate the spectra of nuclei belonging to the group of medium-heavy nuclei, since the boundary between the lightest and medium-heavy nuclei remains unknown. However, the $S_{\text{min-max}}$ method permits estimating the exponents of the spectra and irregularities in the spectra of the lightest and heaviest groups of nuclei in the PCR mass composition that are detected in experiments.

The algorithm of formation of statistical samples of features of EAS initiated by the lightest and heaviest nuclei in the PCR mass composition is as follows:

1. The statistical sample of EAS features that was accumulated in the KASCADE-Grande experiment and which is used in the calculations includes about 100 million events. As the number of experimentally detected events grows, the accuracy of the $S_{\text{min-max}}$ method for separating the groups of the lightest and heaviest nuclei becomes higher.
2. The statistical sample of EAS features that is obtained in the experiment is sorted in an EAS parameter that is sensitive to the type of the PCR nucleus that initiated EAS. Within the $S_{\text{min-max}}$ method, the EAS age is taken to be the parameter that is sensitive to the type of the PCR nucleus.
3. A fixed number of events that corresponds to 1% of events in the total statistics of EAS features is selected. Approximately one million events carrying information about features of EAS initiated by the lightest nuclei in the PCR mass composition upon sorting data in the order of increasing EAS age $S$ and approximately one million events carrying information about features of EAS initiated by the heaviest nuclei in the PCR mass composition upon sorting data in the order of decreasing $S$ were analyzed.

The samples obtained in this way from the KASCADE-Grande database contain the energies $E_0$ of nuclei and other EAS features referring to the groups of the lightest and heaviest nuclei in the PCR mass composition. This made it possible to estimate the exponents of the spectra for groups of nuclei and to reveal irregularities in these spectra. The results of applying the $S_{\text{min-max}}$ method to processing data of the KASCADE-Grande experiment are presented in Fig. 1 in the form of the spectra of the lightest, the heaviest, and all nuclei in the PCR mass composition at energies in the range of $E_0 = 1\text{-}100$ PeV and in Fig. 2 in the form of the spectra of groups of nuclei with the irregularities and exponents of the spectra.

The application of the $S_{min-max}$ method to the database of the KASCADE-Grande experiment permitted obtaining statistical samples of EAS parameters referring, with a highest probability, to EAS initiated by the lightest (heaviest) nuclei in the PCR mass composition.

3. Spectra of the groups of the lightest and heaviest nuclei in the PCR mass composition

The spectra obtained from the KASCADE-Grande statistics of EAS features for the groups of the lightest and heaviest nuclei in the PCR mass composition at energies in the range of $E_0$ = 1-100 PeV are shown in Figs. 1 and 2. The spectra in question were constructed by the $S_{min-max}$ method with the aim of assessing the irregularities and exponents of the spectra.

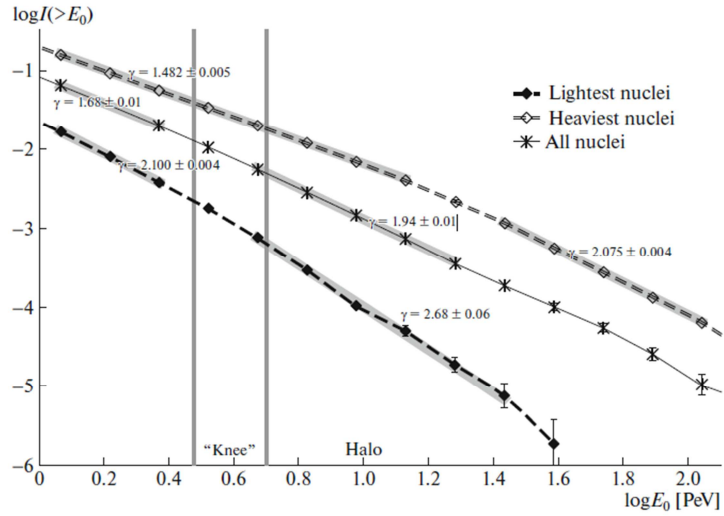

Fig. 1. Integrated $E_0$ spectra of the lightest and heaviest nuclei in the PCR mass composition at energies in the range of $E_0$ = 1-100 PeV, normalization to unity being performed at $E_0$ = 0.1 PeV.

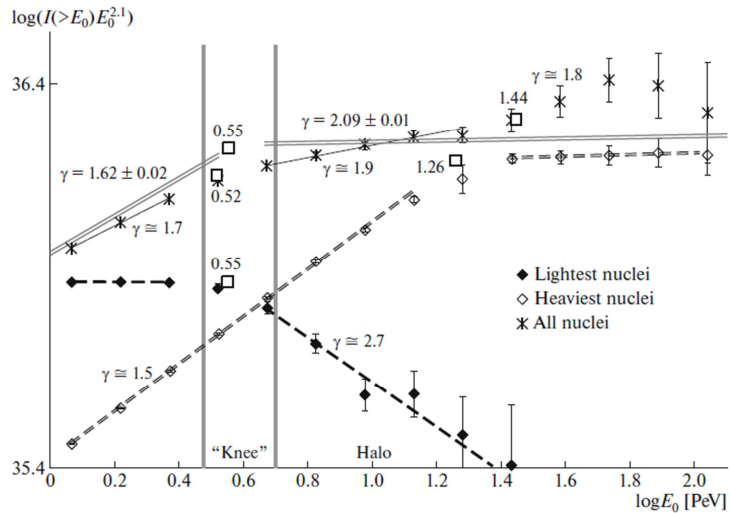

Fig. 2. Integrated spectra of the lightest and heaviest nuclei in the PCR mass composition at energies in the range of $E_0$ = 1-100 PeV without an absolute normalization. The $E_0$ spectrum of EAS that is averaged over the whole statistics of features of experimental data from the KASCADE-Grande experiment is represented by a double gray line with the exponent values of $\gamma$ = 1.62 (before the knee in the spectrum of primary cosmic radiation in the range of $E_0$ = 3-5 PeV) and $\gamma$ = 2.09 (after the knee). The regressions constructed for each group of nuclei by the maximum-likelihood-method are represented by the dashed and solid lines with the points of intersection localizing the knees in the spectra of nuclei. The most probable energy of halo formation in EAS cores [3] at $E_0$ = 10 PeV is indicated.

In Fig. 1, the exponents $\gamma$ characterizing the slopes of the spectra are given for the lightest, heaviest, and mixed groups of nuclei in the integrated $E_0$ spectra of primary cosmic radiation for the statistical sample of about one million events for each group of nuclei. The respective regressions for estimating $\gamma$ were performed by the maximum-likelihood method over data ranges shaded in gray in

Fig. 1. The knees in the spectra of groups of nuclei were determined as the points of intersection of the regression straight lines before and after the presumed localization of a knee in $E_0$ (see Table 2). The data in Table 2 show that the changes in the slopes of the spectra, $\Delta\gamma$, for the lightest and heaviest groups are close, taking a value of $\Delta\gamma \cong 0.6$, whereas $\Delta\gamma$ for the mixed group in the $E_0$ range indicated in Fig. 1 is smaller, $\Delta\gamma \cong 0.2$-$0.3$, which suggests the presence of an additional source of nuclei of the He-CNO group in the knee region of the $E_0$ spectrum of primary cosmic radiation. As is shown in Fig. 2 below, the spectra of the lightest and heaviest nuclei in Fig. 1 do not exhibit a gradual steepening or a gradual growth of the exponent $\gamma$ of the spectrum in $E_0$ but feature a manifest knees (irregularities) of the $E_0$ spectra.

Table 2. Energies at which irregularities in the $E_0$ spectra of nuclei from the PCR mass composition are observed.

| Group of PCR nuclei detected in the experiment | Statistics of first million events from the KASCADE-Grande database | Range of $E_0$, PeV | Slope $\gamma$ of integrated $E_0$ spectrum | Knees in $E_0$ spectra of PCR, PeV | |
|---|---|---|---|---|---|
| Lightest nuclei | Database in the order of increasing S: S = 0.10-0.51 | 1.2-2.3 | 2.100 ± 0.004 | 3.5 | |
| | | 4.7-27.0 | 2.68 ± 0.06 | | |
| All nuclei | S = 0.10-1.48 | 1.2-2.3 | 1.68 ± 0.01 | 3.3 | - |
| | | 4.7-19.1 | 1.94 ± 0.01 | | 27.8 |
| | | 27.0-54.5 | 1.764 ± 0.007 | - | |
| | | 0.1-3 | 1.62 ± 0.02 | 3.6 | |
| | | 5-1000 | 2.09 ± 0.01 | | |
| Heaviest nuclei | Database in the order of decreasing S: S = 1.36-1.48 | 1.2-13.4 | 1.482 ± 0.005 | 18.1 | |
| | | 27.0-109.6 | 2.075 ± 0.004 | | |

In Fig. 2, the spectra of the lightest and heaviest nuclei in the PCR mass composition are given along with details of the change in the exponents $\gamma$ over the range of $E_0 = 1$-$100$ PeV. Figure 2 shows the character of the spectra of the lightest and heaviest nuclei in the region of the knee in the $E_0$ spectrum of primary cosmic radiation, in the region of halo formation in EAS cores at $E_0 = 10$ PeV, and for $E_0$ values around 100 PeV. It is shown by the halo method [1] that, at $E_0 = 10$ PeV, the minimum fraction of $p$+He light nuclei in the PCR mass composition is about 40%, the fractions of protons and helium nuclei being approximately equal to each other. The processing of data from the Pamir experiment shows that, in the case of a smaller fraction of protons in the PCR mass composition, no events induced by gamma-ray families with a halo would appear in experiments employing X-ray emulsion chambers.

The exponents of the spectra of the lightest and heaviest groups of primary cosmic radiation in Fig. 2 are summarized in Table 2. The data in Fig. 2 and in Table 2 show that the knee in the $E_0$ spectrum of the lightest nuclei in the PCR mass composition arises at $E_0 = 10^{0.55}$ PeV $= 3.5$ PeV, which corresponds to a knee in the total spectrum constructed for primary cosmic radiation on the basis of the whole EAS statistics in the KASCADE-Grande experiment. It follows that PCR nuclei before the knee at $E_0 = 3$-$5$ PeV are predominantly light and that the position of the knee in the spectrum of these nuclei determines the knee in the spectrum of all nuclei. The knee in the spectrum of the heaviest nuclei in the PCR mass composition that are detected in the KASCADE-Grande experiment arises at $E_0 = 10^{1.26} = 18$ PeV. The knees for nuclei belonging to other groups lie in the range of $E_0 = 4$-$18$ PeV.

From Fig. 2 and from Table 2, it follows that, in the range of $E_0 = 1$-$100$ PeV, the exponent characterizing the slope of the spectrum of the lightest nuclei, $\gamma \cong 2.1$-$2.7$, is substantially greater than the exponent of the spectrum of the heaviest nuclei, $\gamma \cong 1.5$-$2.1$. This indicates that the heavier fraction of primary cosmic radiation becomes greater.

The data in Fig. 2 indicate that the $E_0$ spectra of PCR nuclei exhibit knees both for the lightest and for the heaviest nuclei in the PCR mass composition; that is, the mechanism of formation of knees in the $E_0$ spectrum of primary cosmic radiation remains invariable for all groups of nuclei. At the same time, the spectrum of all nuclei shows irregularities determining changes of ±0.1 in the exponent of the spectrum in narrow intervals of $E_0$. In particular, the exponent of the spectrum of all nuclei is $\gamma \cong 1.7$ for $E_0 < 3$ PeV, $\gamma \cong 1.9$ for $E_0 = 3$-$27$ PeV, and $\gamma \cong 1.8$ for $E_0 > 27$ PeV. These irregularities in the $E_0$ spectrum of primary cosmic radiation may be due to the procedure used in the experiment or may be indicative of the presence of local sources of nuclei belonging to different groups of primary cosmic radiation.

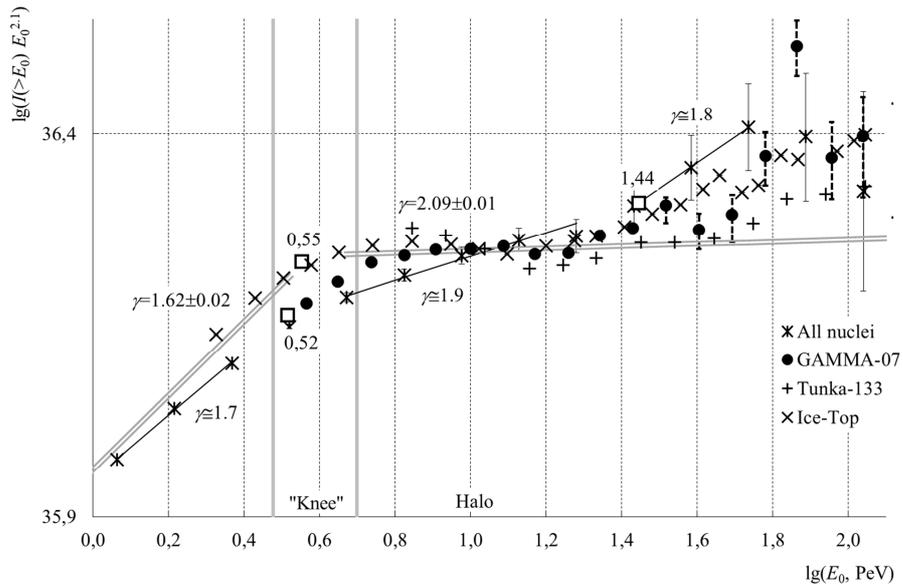

Fig. 3. Integrated $E_0$ spectra obtained in the KASCADE-Grande ("All nuclei") [4] and GAMMA ("GAMMA-07") [11] experiments without an absolute normalization. The maxima of the bumps in the spectra measured in the KASCADE-Grande and GAMMA experiments lie in the range of $E_0$ = 50-75 PeV, which is above the energy of the knee in the spectrum of the heaviest nuclei in the PCR mass composition.

Figure 3 shows that, after the knee in the $E_0$ spectrum of the heaviest PCR nuclei, a bump extending from $E_0 = 10^{1.44} = 28$ PeV to $E_0 \gtrsim 100$ PeV appears in the spectrum of all nuclei. Most probably, $E_0 = 10^{1.74} = 55$ PeV corresponds to this irregularity in the $E_0$ spectrum of primary cosmic radiation for $E_0 > 28$ PeV. A similar irregularity in the $E_0$ spectrum at the most probable value of $E_0 = 10^{1.86} = 72$ PeV was obtained at the GAMMA facility upon processing the GAMMA-07 database (Armenia, Aragats, 700 g/cm$^2$ of the standard atmosphere) [11]. The bump in the $E_0$ spectrum begins after the knee in the spectrum of the heaviest nuclei. This means that the bump cannot be due to the knees in the spectra of PCR nuclei from protons to iron nuclei. Most likely, the bump is formed by an additional source of nuclei of the He-CNO group, since the nature of the bump is not associated with either the lightest or the heaviest nuclei in the PCR mass composition – at the $E_0$ values being considered, the fraction of the lightest nuclei in the PCR mass composition is small; at the same time, the exponent of the spectrum of the heaviest nuclei, $\gamma$, does not change in the range of $E_0$ = 28-100 PeV.

4. Conclusions

The $S_{min-max}$ method has been applied in analyzing data that contain a statistically significant sample of EAS features obtained in the KASCADE-Grande experiment. Knees in the $E_0$ spectra of the groups of the lightest and heaviest nuclei in primary cosmic radiation that are detected in the KASCADE-Grande experiment have been found, and irregularities in the spectra of these groups of nuclei have been analyzed. The following conclusions have been drawn from the resulting spectra:

- A knee in the $E_0$ spectrum of the lightest nuclei in the PCR mass composition is observed at $E_0$ = 3.5 PeV, and the first knee in the $E_0$ spectrum of the mixed mass composition of primary cosmic radiation is associated with the elimination of the lightest nuclei detected in the experiment.
- In the range of $E_0$ = 28-100 PeV, there is a bump in the $E_0$ spectrum of the mixed mass composition of primary cosmic radiation. This bump peaks in the range of $E_0$ = 50-75 PeV and owes its existence to an additional source of PCR nuclei belonging to the He-CNO group.
- In the range of $E_0$ = 1-100 PeV, the fraction of heavy nuclei in the PCR mass composition grows.